\documentclass{article}
\usepackage{spconf,amsmath,graphicx}
\usepackage[skip=0.333\baselineskip]{caption}
\usepackage{subcaption}
\usepackage{multirow}
\usepackage{graphicx}
\usepackage{booktabs}
\usepackage{relsize}
\usepackage{gensymb}
\usepackage{spconf}
\usepackage{MnSymbol}
\usepackage{algorithm}
\usepackage{algpseudocode}
\usepackage{hyperref}
 \usepackage{booktabs}
 \usepackage{multirow}
 \usepackage{graphicx}
 \usepackage[table,xcdraw]{xcolor}

 \algnewcommand\algorithmicforeach{\textbf{for each}}
\algdef{S}[FOR]{ForEach}[1]{\algorithmicforeach\ #1\ \algorithmicdo}

\usepackage[table,xcdraw]{xcolor}

\title{A Dictionary based approach for removing out-of-focus blur}

\name{Uditangshu Aurangabadkar$^\star$, Anil Kokaram$^\dagger$}
\address{Sigmedia Group, Department of Electronic and Electrical Engineering\\
        Trinity College Dublin, Ireland\\
        $^\star$aurangau@tcd.ie, $^\dagger$anil.kokaram@tcd.ie \\
        \textit{www.sigmedia.tv}}
\begin{document}
%

\maketitle
\begin{abstract}
The field of image deblurring has seen tremendous progress with the rise of deep learning models. These models, albeit efficient, are computationally expensive and energy consuming. Dictionary based learning approaches have shown promising results in image denoising and Single Image Super-Resolution. We propose an extension of the Rapid and Accurate Image Super-Resolution (RAISR) algorithm introduced by Isidoro, Romano and Milanfar for the task of out-of-focus blur removal. We define a sharpness quality measure which aligns well with the perceptual quality of an image. A metric based blending strategy based on asset allocation management is also proposed. Our method demonstrates an average increase of approximately 13\% (PSNR) and 10\% (SSIM) compared to popular deblurring methods. Furthermore, our blending scheme curtails ringing artefacts post restoration. 
\end{abstract}
\section{Introduction}
\label{sec:introduction}
In the process of image deblurring (also known as image restoration) we are required to estimate a close approximation ${\hat I}$, of the original image $I$ from a blurred image $G$. Classically the degradation is modelled as follows:
\begin{equation}
    G = I * H_K + \epsilon
\end{equation}
where $H_K$ is a blur kernel of size $K$, and $\epsilon \in {\cal N}(\sigma_e^2, 0)$ is additive noise. 
Depending on whether the blur kernel is known or unknown the restoration process is called \textbf{non-blind} or \textbf{blind} respectively. In this work we are concerned with \textbf{non-blind} restoration and optical blur only, as there have been relatively few published papers addressing the problem of optical blur removal \cite{karaali2022svbr}. This type of degradation is important in the case of pre-computed camera correction to improve the baseline image quality, especially for high-quality apparatus~\cite{rosney2022automating}. 

Classical methods for restoration such as the Wiener filter and iterative techniques like Richardson-Lucy (R-L) deconvolution \cite{lucy1974iterative} have recently given way to the use of DNNs \cite{ronneberger2015u, quan2023single} producing state-of-the-art results. The drawback of these architectures, however, is their large demand on energy and resources. Competing processes have recently emerged by advancing iterative approaches \cite{belyaev2022black}.
Several works \cite{haywood2023real, lopez2023deep} 
have combined traditional methods with DNN architectures to build inexpensive and energy efficient models. In \cite{lopez2023deep} for example, the Moore-Penrose pseudoinverse is combined with a low-complexity DNN for the task of blur artefact removal.

The intuitive advance gained from DNN based models is their ability to inject hard-to-model knowledge about sharp images into a restoration process. Therefore, in this work we adopt a classical strategy initially proposed by Romano, Isidoro, Milanfar et al. \cite{romano2016raisr, bladeMilanfar} for filter learning and adapt it for blur removal. 

Two other issues arise in the task of blur removal. One is that the loss functions or quality metrics which are typically deployed (e.g. PSNR, SSIM) do not exactly target the perception of image sharpness. The other is that ringing artefacts remain difficult to suppress successfully. In this paper we adapt metric $Q$~\cite{zhu2010automatic} which is directly related to image sharpness, and further deploy it in a novel strategy for blending restored image candidates in an effort to address the ringing problem. Image blending has been utilized to produce better quality images, an example of which is the CT-based blending scheme \cite{zabih1994non}. This blending scheme was employed in RAISR - a Super-Resolution algorithm, to blend two images by selecting the most desirable characteristics depending on the structure of the images. Inspired by this, our blending strategy employs a little known non-parametric approach first developed by the Nobel laureate Harry Markowitz \footnote{https://www.nobelprize.org/prizes/economic-sciences/1990/markowitz/facts/}. As opposed to CT-based blending, our mechanism is capable of blending several images, instead of two. 

The novelty introduced in this paper is therefore summarized as follows:
\begin{itemize}
    \item The development of a dictionary based out-of-focus blur removal mechanism dependent on patch eigenanalysis. 
    \item A novel image blending mechanism based on a no-reference metric $Q$.
    \item A novel approach to characterise sharpness improvement in deblurring by developing a bounded adaptation of $Q$. 
\end{itemize}

\subsection{Background}
\label{sec:background}
Dictionary based image models have been used for image enhancement for some time \cite{routray2019efficient}. The general idea is to associate observed patches with a dictionary of prototypical image patches learned from a corpus of training data. These are associated with some sort of inverse filter which produces the enhancement effect.  

Rapid and Accurate Image Super-Resolution (RAISR) is a dictionary based learning model introduced for the task of Single Image Super-Resolution. Properties based on the eigenanalysis of image patches are learned between a pair of high-resolution and low-resolution images and stored in a hash \textit{table}, with the estimated filter being the hash \textit{key}.  The BLADE algorithm (Best Linear Adaptive Enhancement) \cite{bladeMilanfar} was introduced as an extension of RAISR that provides a framework for filter learning and visualisation. In the work published by Choi et al. \cite{choi2018fast}, such schemes have produced state-of-the-art results in the domain of denoising. 

\subsection{Learning Filters}
\label{sec:filter_learning}
The core idea of the dictionary based deblurring methods is to learn an optimal deblurring filter for all possible image patches. A dictionary of patch textures is parameterised with texture eigenvalues and associated with corresponding filters in a Lookup Table (LUT).

Given a degraded image decomposed into patches of size $k \times k$, the goal is to estimate coefficients of a filter $\textbf{h}$, such that the deconvolution of a degraded patch with that filter results in a patch \textit{similar} to the one in the original image. The optimisation of ${\bf h}$ is expressed as follows. 
\begin{equation}
    \min_h \mathlarger{\sum_{i=1}^{N}} ||\textbf{A}_i\textbf{h} - \textbf{b}_i||_2^{2}
    \label{eq:min_prob}
\end{equation}
where $\textbf{A}$ denotes a patch of size $k \times k$ from the degraded image, $\bf{h}$ denotes the filter and $\textbf{b}$ denotes the corresponding clean, undegraded patch. Following the development in RAISR, where the authors argue that storing matrix $\textbf{A}$ in memory is a computationally expensive and inefficient task, the problem is redefined as follows:
\begin{equation}
    \min_h ||\textbf{W}\textbf{h}-\textbf{V}||_2^{2}
    \label{eq:red_min}
\end{equation}
where $W$ = $\textbf{A}^T$\textbf{A} and $V$ = $\textbf{A}^T$\textbf{b}. Filter indexing is achieved by measuring three properties of each patch in the degraded image: angle $\theta$, strength $\sqrt{\lambda_1}$ and coherence $\mu$. To do this, the gradient at each pixel is measured, producing matrix \textbf{G}, followed by the calculation of eigenvalues and eigenvectors of the matrix $\textbf{G}^T\textbf{G}$. Coherence ($\mu$) characterizes the local anisotropy of a patch \cite{choi2018fast} and is given by:
\begin{equation}
    \mu = \frac{(\sqrt{\lambda_1} - \sqrt{\lambda_2})}{(\sqrt{\lambda_1} + \sqrt{\lambda_2})}
\end{equation}
where $\sqrt{\lambda_1}$ and $\sqrt{\lambda_2}$ are the maximum and minimum eigenvalues respectively. The angle ($\theta$) is given by:
\begin{equation}
    \theta = \arctan(\phi_{max}^{x}, \phi_{max}^{y})
\end{equation}
where $\phi_{max}$ denotes the eigenvector corresponding to the largest eigenvalue of a patch. The strength is given by the square root of the largest eigenvalue -- $\sqrt{\lambda_1}$ . In our implementation, these parameters are then quantized using step sizes 24, 3, 3 respectively.  The dictionary is finally extended using patch rotations of 45$^\circ$ and 90$^\circ$ as well as reflections on each axis. 

In inference, given a degraded patch, we measure the corresponding texture vector $(\theta, \mu, \sqrt{\lambda_1})$. The quantized vector is then used to look up the relevant restoration filter for that patch in our dictionary. That filter is then applied to the patch to restore it. The process is repeated for all patches. Please see \cite{romano2016raisr} for more details as well as the supplementary material\footnote{https://github.com/aurangau/ICIP2024}.

\section{Adaptation to Deblurring}
\label{sec:model}
  For deblurring, we adapt the learning approach by utilizing pairs of sharp--blurry images for estimating filter coefficients. In our scenario, given a blurry image, we wish to apply a set of filters to each patch, such that deconvolved patch is a \textit{similar} match to corresponding patch in the sharp image.
 To populate our filterbank, we utilised 500 images from the BSD500~\cite{martin2001database} dataset. As part of the training, two different blur types were used -- Gaussian blurs and box blurs. For the experiments (see section~\ref{sec:res}), Gaussian blur kernels of sizes $11 \times 11$, $\sigma=1.85$ and $15 \times 15$, $\sigma=2.10$ were utilised. In addition to this, box blur of size $3 \times 3$ was also used. Rather than deploy an iterative gradient based scheme (as in RAISR) for solving the core least squares problem, we deploy the Moore-Penrose pseudoinverse. This pseudoinverse  produced sharper images with higher metric values (see supplementary material). Another key difference between RAISR and our adaptation is the omission of an intermediate restoration step, which in SR is bilinear upsampling. 
 Finally, we deploy a different blending strategy for combining different restored versions of the image. This is described in section~\ref{sec:index_j_blending}. 

\subsection{Sharpness Quality Measure}
Full-reference metrics such as PSNR or SSIM have been largely used for determining the perceptual quality of an image. However, such metrics, albeit simple to interpret, do not provide detail about the sharpness of an image.
Metric $Q$, as stated in section~\ref{sec:introduction}, is a no-reference metric that utilizes image gradient statistics to provide an indication of the sharpness of an image. For computing $Q$, an image is broken into $N$ non-overlapping patches, followed by eigendecomposition, which produces the singular values $s_1$ and $s_2$ for each patch. Local coherence $R$, based on the difference between the eigenvalues is calculated, followed by an application of threshold $\tau$ for determining the anisotropy of a patch. Finally, $Q_k$ (where $k$ denotes the corresponding patch) is calculated for each anisotropic patch, given by: 
\begin{equation}
    Q_k = s_1 * \frac{s_1 - s_2}{s_1 + s_2}; \hspace{1.5mm} Q = \mathlarger{\sum}_{k=1}^{N} Q_k
    \label{eq:loc_metric}
\end{equation}
As $Q$ measures the strong directionality of the dominant eigenvalues, a higher value correlates to a sharper image. Hence it is useful as a measure of blur and noise. However, this also means that it remains susceptible to ringing artefacts, in the sense that an image with these artefacts can yield a larger value of Q than in the original image. We show an example of this in the supplementary material\footnote{{https://github.com/aurangau/ICIP2024}}. Furthermore as $Q$ is unbounded (it is also a no-reference metric), it is difficult to associate measured values of $Q$ to relative changes in image quality. We therefore design a metric $J$ based on $Q$ that is bounded within the interval $[0, 1]$.

Consider a pair of undegraded--degraded images $I$ and $G$, and a restored image $\hat{I}$ as defined previously.  Our bounded measure $J(\hat{I}, I)$ needs to be defined such that if $\hat{I}=G$ then $J(\hat{I}, I) = 0$ and when $\hat{I}=I$, then $J(\hat{I}, I)=1$. 
Assuming that $Q_{\hat{I}}$ is \textit{well-behaved}, i.e. $Q_{\hat{I}} \in [Q_G, Q_I]$, we first measure the relative position in $Q$-space of $\hat{I}$ on the continuum between $G$ and $I$ by defining a deviation $V(\cdot, \cdot)$ which is as follows:
\begin{equation}
   V = \left | \frac{Q_{\hat{I}} - Q_I}{Q_{\hat{I}} - Q_G} \right |
    \label{eq:dev_ratio}
\end{equation}
Our bounded, reference sharpness measure $J$ is now defined as follows:
\begin{equation}
    J=
       \frac{1}{1 + V}
\end{equation}
Hence when $\hat{I} = I$, $V = 0, J = 1$ and when $\hat{I} = G$, $V = \infty, J = 0$. 

\subsection{Metric $Q$ Guided Blending Strategy}
\label{sec:index_j_blending}
As discussed in section~\ref{sec:introduction}, to reduce ringing we create a set of images restored using different patch sizes which exhibit different levels of ringing. We wish to devise a blending scheme to combine the best features across these images, thereby producing an image with reduced ringing. An analogous problem, in Modern Portfolio Theory (MPT), is that of asset allocation, where the goal of any investor is to maximize the returns on investments and simultaneously reduce the risk attached to each asset by manipulating the weights attributed to corresponding assets. A good example of such a strategy is the Markowitz model~\cite{markowitz52}. We develop our image blending scheme using this idea. 

In our task, we assume the values of pixels in a blended image to be the expected return of the \textit{portfolio}. The value of $Q$ is the \textit{return} from each \textit{asset}. An asset, in our task, is an image. A higher value of $Q$ also corresponds to a higher \textit{risk} associated with the respective asset, which in our case, is ringing in the respective image. Therefore, our problem can now be modified as maximizing the overall \textit{return} ($Q$) of the \textit{portfolio} (blended image), while simultaneously minimizing the \textit{risk} (ringing) associated with each \textit{asset} (image).

We create a set of restored images $\hat{I} = \{\hat{I}_1, \hat{I}_2, \ldots, \hat{I}_N\}$, by modulating the patch of size $p \times p$ during inference. For each image, we calculate metric $Q$ denoted by $Q_1$, $Q_2, \ldots, Q_N$, which we arrange in an ascending order of their magnitudes.  Our blending strategy follows an iterative update to the weighting coefficients $w_i^{m}$, $i, m \in [0, N-1]$. Each application of the blending algorithm on an image can be termed as a \textit{round}.  We employ several rounds of our iterative method to the weights $w_i^{m}$, $i, m \in [0, N-1]$. We initialize our algorithm by applying equal weighting to all images $\hat{I}_j$, $j \in [1, N]$, i.e. $w_i^{0} = \frac{1}{N}$: 

\begin{equation}
   {I}_{B}^{0} = w_0^{0}\hat{I}_1 + w_1^{0}\hat{I}_2 + \ldots + w_{N-1}^{0}\hat{I}_N
    \label{eq:equal_weights}
\end{equation}
 At each iteration the key intuition is to reduce the value of weights applied to lower quality assets (blurry restorations) and increase proportionally the weights applied to higher value assets (sharper restorations). Given a pair of images $I_p, I_q$ the change in weight is denoted as $\delta_{\{I_p, I_q\}}$,
 \begin{equation}
     \delta_{\{I_p, I_q\}} = \frac{Q_{I_q} - Q_{I_p}}{Q_{I_p}}
 \end{equation}
In the first iteration of each round, we compare the image with the largest $Q$, $\hat{I}_N$, with each of the other restored versions and increase the value of $w_{N-1}$ by $\delta$ while reducing the value of $w_i, i \in [0, N-2]$ by the same amount. This ensures that the sum of the weights is always unity. We repeat this for each of the largest weights as detailed in Algorithm 1.
\begin{algorithm}
\caption{Metric $Q$ Guided Blending}
\begin{algorithmic}[1]
    \State \textbf{Input:} Images $\hat{I}_1, \hat{I}_2, \ldots \hat{I}_{N}$
    \State \textbf{Output:} Blended Image $I_B$
    \State Measure [$Q_1, Q_2, \cdots, Q_N$] = $\textbf{Q}(\hat{I}_1, \hat{I}_2, \cdots, \hat{I}_N)$; 
    \State Order $\hat{I}_u$, $u \in [1, N]$ such that $Q_1 < Q_2 < \cdots < Q_N$; 
    \State Initialize $w_k^{0} = \frac{1}{N}$, $k \in [0, N-1]$; 
    \State $I_{B}^0$ = $w_0^{0}\hat{I}_1 + w_1^{0}\hat{I}_2 + \cdots w_{N-1}^{0}\hat{I}_N$;
    \ForEach{iteration $i$ = 1 to $(N-1)$}
        
        \For{$k = 0$ to $N-i-1$}
            \State $w_{k}^{i}$ = $w_{k}^{i-1}$ - $\delta_{\{k+1, N\}}^{i}$
            \State $w_{N-i}^{i}$ = $w_{N-1}^{i-1}$ + $\mathlarger{\sum_{j=1}^{N-i-1}}\delta_{\{j, N\}}^{i}$
        \EndFor
    \EndFor
    
    \State $I_B$ = ${w}_0^{N-1}\hat{I}_1 + {w}_1^{N-1}\hat{I}_2 + \ldots + {w}_{N-1}^{N-1}\hat{I}_N$ 
\label{alg:Q_blending}
\end{algorithmic}
\end{algorithm}

As the restored images $\hat{I}_j$, $j \in [1, N]$ do not change between iterations, $\delta$ remains constant across rounds for each weighting coefficient. Hence the increase in weights forms an Arithmetic Progression after each round. We can therefore measure the exact number of rounds it would take for $w_0$ to become 0, thus giving us a maximum number of rounds that can be applied.

We wish to stop application of rounds when the output blended image no longer increases in sharpness $Q$. Our termination criteria (at round $m$) are therefore as follows: 
\begin{enumerate}
    \item $Q(I_{B}^{m+1}) < Q(I_{B}^{m})$ -- When no further increase is seen in image quality, signalling a monotonic decrease in $Q$.
    \item For a threshold $\eta$, $Q(I_B^{m+1}) - Q(I_B^{m}) \geq \eta$.
    \item $w_{0}^{m} \approx 0$ -- We no longer use all images for blending.
\end{enumerate}
\section{Results}
\label{sec:res}
\noindent{\textbf{Non-Blind Blur Removal}}:
We evaluate the impact of two types of blur kernels: Gaussian blur of size 15 $\times$ 15, $\sigma$ = 2.10 and a box blur of size 3 $\times$ 3. A total of 19 images were used for inference - 14 from set14 \cite{zeyde2012single} and 5 from set5 \cite{bevilacqua2012low}. The baseline patch size chosen for both training and inference was set to 21. Our model was also trained with varying patch sizes, which were utilized for blending as discussed in section~\ref{sec:index_j_blending}. The proposed method was further tested on 800 images from the DIV2K~\cite{Agustsson_2017_CVPR_Workshops} dataset and produces images with a PSNR of approximately 29 dB.

\begin{figure}
    \centering
    
    \subfloat[Original]{\includegraphics[width=0.16\textwidth]{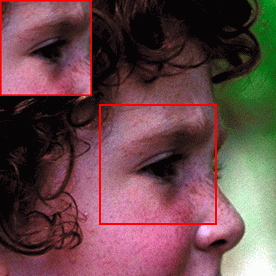}} \hfill
    \subfloat[Degraded]{\includegraphics[width=0.16\textwidth]{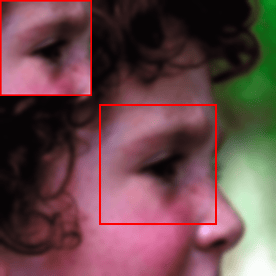}} \hfill
    \subfloat[Restormer \cite{zamir2022restormer} ]{\includegraphics[width=0.16\textwidth]{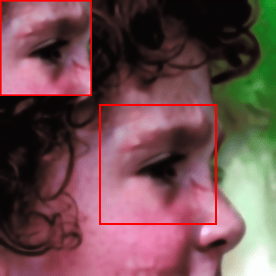}} \\
    
    \subfloat[IFAN \cite{lee2021iterative}]{\includegraphics[width=0.16\textwidth]{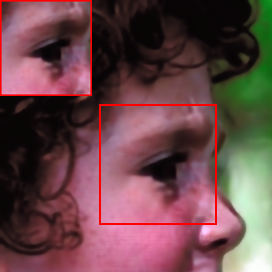}} \hfill
    \subfloat[NBDNet \cite{chen2021learning}]{\includegraphics[width=0.16\textwidth]{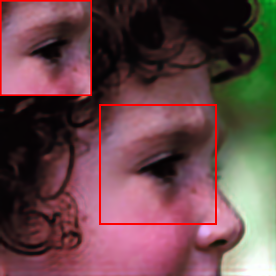}} \hfill
    \subfloat[Landweber \cite{wang2023reverse}]{\includegraphics[width=0.16\textwidth]{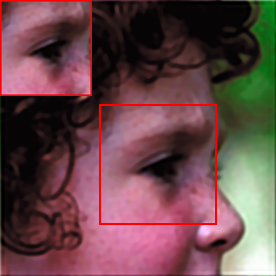}} \\
    
    \subfloat[NA Landweber]{\includegraphics[width=0.16\textwidth]{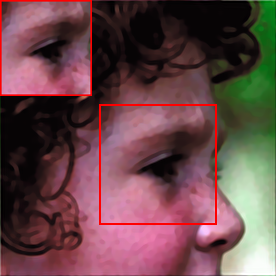}} \hfill
    \subfloat[PC Landweber]{\includegraphics[width=0.16\textwidth]{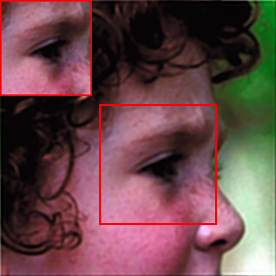}} \hfill
    \subfloat[Ours]{\includegraphics[width=0.16\textwidth]{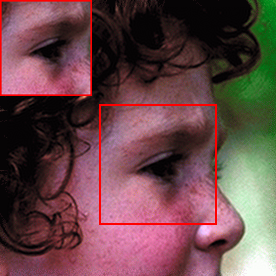}}
    
    \caption{Visual comparison of deblurring techniques. Our technique exhibits better detail preservation and less spurious texture artefacts especially c.f. Restormer, IFAN and NBDNet. }
    \label{fig:all_comp}
\end{figure}

We compare our restoration technique with models designed for the purposes of blind deblurring - Restormer \cite{zamir2022restormer}, IFAN \cite{lee2021iterative}, and non-blind deblurring - DNN based -- NBDNet \cite{chen2021learning}, Landweber-like iterative defiltering \cite{wang2023reverse}, Nesterov Accelerated Landweber-like iterative defiltering (NAL) and Phase Corrected Landweber-like iterative defiltering (PCL). We must highlight that due to a lack of existing DNN methods for removing defocus blur using a known blur kernel, we only compare our model against NBDNet. 
Table~\ref{tab:results_table} reports the PSNR, SSIM and $J$ using these restoration techniques. We also note a 60\% increase w.r.t BRISQUE~\cite{mittal2012no} as compared to methods listed in table~\ref{tab:results_table}. See supplementary materials~\footnote{https://github.com/aurangau/ICIP2024}. The methods highlighted in grey are modifications of the Landweber-like iterative method~\cite{wang2023reverse}. The maximum iterations for the Landweber-like method was set to 500, 125 for NAL and 200 for PCL. In general the Landweber-like techniques take approximately 5 to 10 minutes per $512\times 512$ image, the learned filter techniques take 25 seconds and DNNs about 1 to 2 seconds of compute time. Note that the DNN inference is conducted on GPUs while the other methods are executed on CPUs. All experiments were conducted on an Intel i9-10900F@2.80 GHz CPU with a GeForce RTX2080Ti GPU.
As evident from table~\ref{tab:results_table}, our method compares favourably to those considered to be state-of-the-art. Figure~\ref{fig:all_comp} provides a visual comparison of the various methods tested on the image \textit{Face} (512 $\times$ 512). This comparison shows that our technique exhibits better agreement with the original image including better texture preservation. 

\begin{table}[]
\centering
\begin{tabular}{@{}lcccc@{}}
\toprule
\textbf{Algorithm} & \multicolumn{1}{l}{\textbf{PSNR (dB)}} & \textbf{$J$} & \multicolumn{1}{l}{\textbf{SSIM}} & \multicolumn{1}{l}{\textbf{BRISQUE}} \\ \midrule
IFAN \cite{lee2021iterative}                  & 23.100                   & 0.322                   & 0.653                   & 42.379                   \\
Restormer \cite{zamir2022restormer}              & 24.566                   & 0.403                   & 0.723                   & 44.342                   \\
NBDNet \cite{chen2021learning}                 & 27.280                   & 0.660                   & 0.754                   & 40.291                   \\
Landweber \cite{wang2023reverse}              & 26.263                   & 0.693                   & 0.757                   & 42.017                   \\
\rowcolor[HTML]{C0C0C0} 
NA Landweber           & 26.291                   & 0.770                   & 0.758                   & 40.843                   \\
\rowcolor[HTML]{C0C0C0} 
PC Landweber           & 27.732                   & 0.775                   & 0.820                   & 38.736                   \\
\textit{\textbf{Ours}} & \textit{\textbf{29.287}} & \textit{\textbf{0.893}} & \textit{\textbf{0.847}} & \textit{\textbf{24.670}} \\ \midrule
IFAN                   & 25.409                   & 0.568                   & 0.789                   & 44.813                   \\
Restormer              & 26.645                   & 0.594                   & 0.823                   & 44.642                   \\
NBDNet                 & 31.036                   & 0.742                   & 0.869                   & 41.583                   \\
Landweber              & 29.011                   & 0.730                   & 0.849                   & 42.649                   \\
\rowcolor[HTML]{C0C0C0} 
NA Landweber           & 30.019                   & 0.800                   & 0.867                   & 41.444                   \\
\rowcolor[HTML]{C0C0C0} 
PC Landweber           & 30.715                   & 0.816                   & 0.890                   & 39.477                   \\
\textit{\textbf{Ours}} & \textit{\textbf{31.999}} & \textit{\textbf{0.891}} & \textit{\textbf{0.900}} & \textit{\textbf{26.814}} \\ \bottomrule
\end{tabular}
\caption{Comparison of deblurring techniques using PSNR, SSIM and our new metric $J$. (Top) Set14. (Bottom) Set5. N.B. $J$ is better behaved than SSIM as it expresses over a wider numerical range (0.3--0.9) than SSIM c.f. (0.7--0.9)}
\label{tab:results_table}
\end{table}

\begin{figure}
    \begin{subfigure}{0.480\linewidth}
      \includegraphics[width=\linewidth]{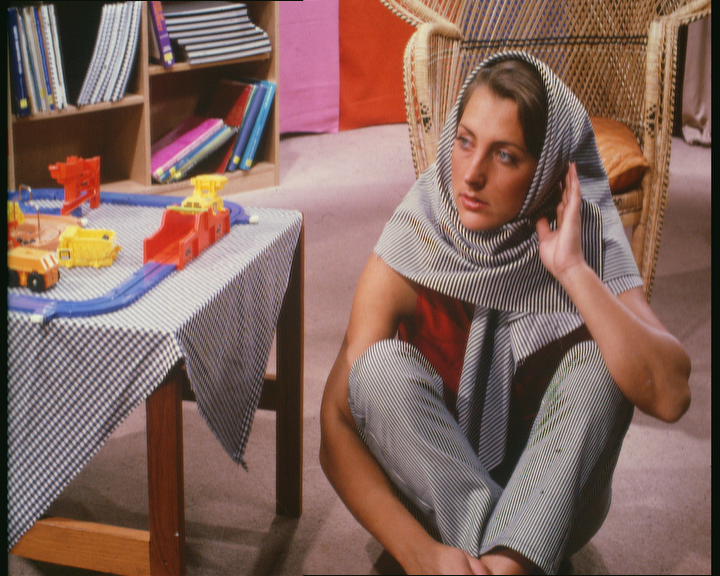}
      \caption{$J$ = 1.0, $Q$ = 11.901}
      \label{fig:original_barbara}
    \end{subfigure}\hfill 
    \begin{subfigure}{0.480\linewidth} 
      \includegraphics[width=\linewidth]{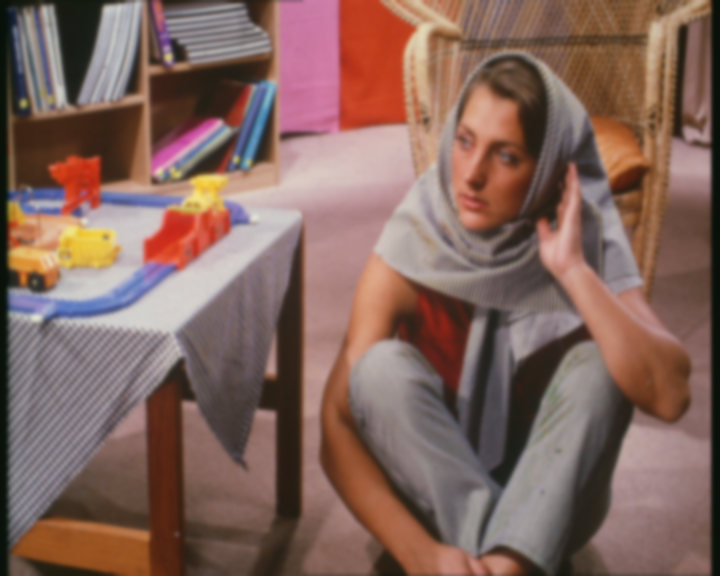}
      \caption{ $J$ = 0.0, $Q$ = 5.638}
      \label{fig:blurry_barbara}
    \end{subfigure}
    
    \medskip 
    \begin{subfigure}{0.480\linewidth}
      \includegraphics[width=\linewidth]{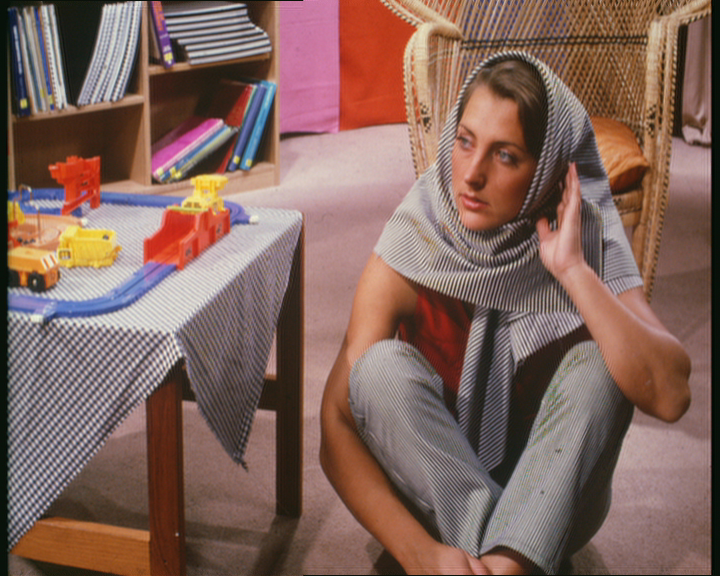}
      \caption{$J$ = 0.703, $Q$ = 10.055}
      \label{fig:p13_barbara}
    \end{subfigure}\hfill 
    \begin{subfigure}{0.480\linewidth}
      \includegraphics[width=\linewidth]{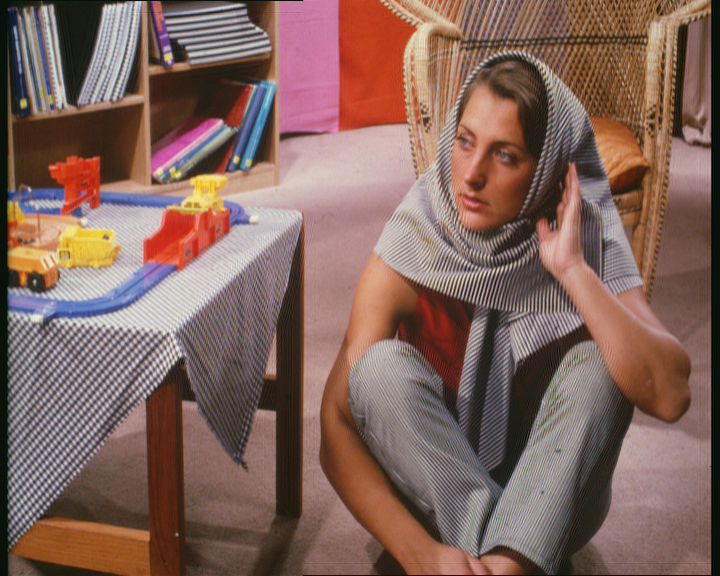}
      \caption{$J$ = 0.779, $Q$ = 10.064}
      \label{fig:p21_barbara}
    \end{subfigure}

    \caption{Behaviour of index $J$. Top Row: Original Image (Left),  Degraded (blurry) Image (Right). Bottom Row: $\hat{I}$, $p$ = 13 (Left), $\hat{I}$, $p$ = 21. $J$ better reflects changes in sharpness while $Q$ is almost identical for the two restored versions.}
    \label{fig:roc}
\end{figure}

\noindent{\textbf{Index $J$} \label{sec:index_J}}:
To better understand the correlation between index $J$ and picture quality, Fig.~\ref{fig:blurry_barbara} shows image \textit{Barbara} (576 $\times$ 720) degraded  with a Gaussian blur kernel 15 $\times$ 15. Restorations using our method with two patch sizes; 13 (fig.~\ref{fig:p13_barbara}) and 21 (fig~\ref{fig:p21_barbara}) are also shown. A larger $Q$ or $J$ always indicates a sharper image e.g. the original image and the restored versions all have higher values of $Q$ and $J$ than the blurry image. However, there is a relatively small change in $Q$  for the two restored images, $Q=10.05,10.06$ respectively. In contrast, there is a noticeable difference in index $J$ values  i.e. $J=0.7,0.78$ respectively. Therefore $J$ separates the different restorations more effectively than just $Q$ in this instance. Note that the image has a well-behaved $Q$, which explains why $J$ correlates more effectively with perceptual quality. 

\noindent{\textbf{Metric $Q$ Guided Blending}: 
Fig.~\ref{fig:blended_im} shows an example result of our blending scheme (section~\ref{sec:index_j_blending}) for reducing ringing.
Four restored versions of the blurry image were generated and sharpness measured with $Q$. Ordering these w.r.t. $Q$, we observed $Q(\hat{I}_{21})<Q(\hat{I}_{25})<Q(\hat{I}_{15})<Q(\hat{I}_{13})$.
Hence the blended estimate $\hat{I}_B^0$ is expressed as follows:
\begin{equation}
    \hat{I}_B^{0} = w_1^0 \cdot \hat{I}_{21} + w_2^0 \cdot \hat{I}_{25} + w_3^0 \cdot \hat{I}_{15} + w_4^0 \cdot \hat{I}_{13}
\end{equation}
Using the algorithm as described previously, we terminate after 2 rounds in this case using round 1 as the output because $Q(\hat{I}_B^2) < Q(\hat{I}_B^1)$. The final blending combination in this case was as follows: 
\begin{equation}
    \hat{I}_B^{2} = 0.2451 \cdot \hat{I}_{21} + 0.2487 \cdot \hat{I}_{25} + 0.2515 \cdot \hat{I}_{15} + 0.2547 \cdot \hat{I}_{13}
\end{equation}
At the end of this blending process $Q(\hat{I}_B^1)=5.278$ which improved the overall score c.f. initial weights ($5.270$) slightly. The impact on ringing though is shown visually in Fig.~\ref{fig:blended_im}. The  restored image $\hat{I}_{13}$ with a patch size of 13 exhibits ringing artefacts on the right edges of the flower petals (highlighted in the red box) while these artefacts are significantly suppressed in the blended image $I_B^{1}$ (fig.~\ref{fig:high_Q_blend}). This method was tested on 300 images and showed a 1.3\% increase in sharpness($Q$) compared with averaging only. For scaling purposes, we multiply $Q$ by a factor of 64. 

\begin{figure}
    
    \subfloat[Original Image. $Q$=5.365, $J$=1.0]{\includegraphics[width=0.14\textwidth]{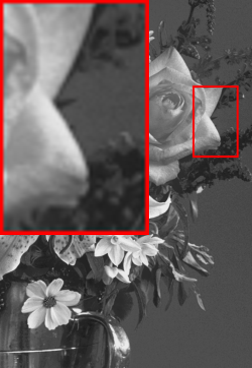}} \hfill
    \subfloat[Image $\hat{I}_{13}$. $Q$=5.275, $J$=0.958]{\includegraphics[width=0.14\textwidth]{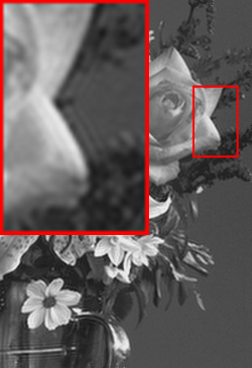}} \hfill
    \subfloat[Image $I_{B}^1$. $Q$=5.278, $J$=0.959]{\includegraphics[width=0.14\textwidth]{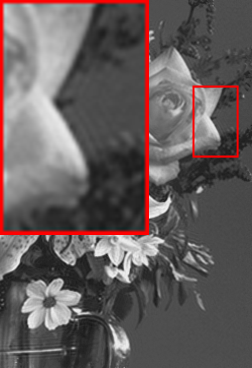}  \label{fig:high_Q_blend}}
    \caption{Comparison of Blended Image with Original and Restored Images}
    \label{fig:blended_im}
\end{figure}

\section{Discussion and Conclusion}
\label{sec:discussion}
We have presented an extension of a dictionary based learning method for a previously unexamined image degradation model -- out-of-focus blur. Our technique performs better than a range of both non-blind and blind restoration techniques across all metrics. Averaging the improvement between our technique and each of the existing 6 techniques tested, we observe an increase of 13\% in PSNR and 10\% in SSIM.
We have also proposed a new sharpness quality measure derived from an existing no-reference metric to quantify qualitative sharpness change w.r.t. original reference image. This new sharpness metric $J$ is better scaled and expresses over a wider range within (0--1) than SSIM for the specific case of sharpness. Drawing inspiration from the principles of dynamic asset allocation and Modern Portfolio Theory (MPT), we have devised a novel image blending mechanism where the weighting coefficients are dependent on the relative sharpness of the candidate restorations. This allows us to produce images with fewer ringing artefacts.
Our future work involves applying this blending strategy \textit{per patch} instead of the entire image.

\clearpage
\bibliographystyle{IEEEbib}
\bibliography{strings,refs}

\begin{thebibliography}{10}

\bibitem{karaali2022svbr}
Ali Karaali and Cl{\'a}udio~Rosito Jung,
\newblock ``Svbr-net: A non-blind spatially varying defocus blur removal network,''
\newblock in {\em 2022 IEEE International Conference on Image Processing (ICIP)}. IEEE, 2022, pp. 566--570.

\bibitem{rosney2022automating}
Sophia Rosney, Ciar{\'a}n Donegan, Meegan Gower, Wissam Jassim, Hugh Denman, Donal Scannell, and Anil Kokaram,
\newblock ``Automating sports broadcasting using ultra-high definition cameras, neural networks, and classical denoising,''
\newblock in {\em Applications of Digital Image Processing XLV}. SPIE, 2022.

\bibitem{lucy1974iterative}
Leon~B Lucy,
\newblock ``An iterative technique for the rectification of observed distributions,''
\newblock {\em Astronomical Journal, Vol. 79, p. 745 (1974)}, vol. 79, pp. 745, 1974.

\bibitem{ronneberger2015u}
Olaf Ronneberger, Philipp Fischer, and Thomas Brox,
\newblock ``U-net: Convolutional networks for biomedical image segmentation,''
\newblock in {\em Medical Image Computing and Computer-Assisted Intervention--MICCAI 2015: 18th International Conference, Munich, Germany, October 5-9, 2015, Proceedings, Part III 18}. Springer, 2015, pp. 234--241.

\bibitem{quan2023single}
Yuhui Quan, Xin Yao, and Hui Ji,
\newblock ``Single image defocus deblurring via implicit neural inverse kernels,''
\newblock in {\em Proceedings of the IEEE/CVF International Conference on Computer Vision}, 2023, pp. 12600--12610.

\bibitem{belyaev2022black}
Alexander~G Belyaev and Pierre-Alain Fayolle,
\newblock ``Black-box image deblurring and defiltering,''
\newblock {\em Signal Processing: Image Communication}, vol. 108, 2022.

\bibitem{haywood2023real}
Charlie Haywood and Rabih Younes,
\newblock ``Real-time blind deblurring based on lightweight deep-wiener-network,''
\newblock in {\em 2023 International Joint Conference on Neural Networks (IJCNN)}. IEEE, 2023, pp. 1--8.

\bibitem{lopez2023deep}
Santiago L{\'o}pez-Tapia, Javier Mateos, Rafael Molina, and Aggelos~K Katsaggelos,
\newblock ``Deep robust image restoration using the moore-penrose blur inverse,''
\newblock in {\em 2023 IEEE International Conference on Image Processing (ICIP)}.

\bibitem{romano2016raisr}
Yaniv Romano, John Isidoro, and Peyman Milanfar,
\newblock ``Raisr: Rapid and accurate image sieeebibuper resolution,''
\newblock {\em IEEE Transactions on Computational Imaging}, vol. 3, no. 1, pp. 110--125, 2016.

\bibitem{bladeMilanfar}
Pascal Getreuer, Ignacio Garcia-Dorado, John Isidoro, Sungjoon Choi, Frank Ong, and Peyman Milanfar,
\newblock ``Blade: Filter learning for general purpose computational photography,''
\newblock in {\em 2018 IEEE International Conference on Computational Photography (ICCP)}, 2018, pp. 1--11.

\bibitem{zhu2010automatic}
Xiang Zhu and Peyman Milanfar,
\newblock ``Automatic parameter selection for denoising algorithms using a no-reference measure of image content,''
\newblock {\em IEEE transactions on image processing}, vol. 19, no. 12, pp. 3116--3132, 2010.

\bibitem{zabih1994non}
Ramin Zabih and John Woodfill,
\newblock ``Non-parametric local transforms for computing visual correspondence,''
\newblock in {\em Computer Vision—ECCV'94: Third European Conference on Computer Vision Stockholm, Sweden, May 2--6 1994 Proceedings, Volume II 3}. Springer.

\bibitem{routray2019efficient}
Sidheswar Routray, Arun~Kumar Ray, and Chandrabhanu Mishra,
\newblock ``An efficient image denoising method based on principal component analysis with learned patch groups,''
\newblock {\em Signal, Image and Video Processing}, 2019.

\bibitem{choi2018fast}
Sungjoon Choi, John Isidoro, Pascal Getreuer, and Peyman Milanfar,
\newblock ``Fast, trainable, multiscale denoising,''
\newblock in {\em 2018 25th IEEE International Conference on Image Processing (ICIP)}. IEEE, 2018, pp. 963--967.

\bibitem{martin2001database}
David Martin, Charless Fowlkes, Doron Tal, and Jitendra Malik,
\newblock ``A database of human segmented natural images and its application to evaluating segmentation algorithms and measuring ecological statistics,''
\newblock in {\em Proceedings Eighth IEEE International Conference on Computer Vision. ICCV 2001}. IEEE, 2001, vol.~2, pp. 416--423.

\bibitem{markowitz52}
Harry Markowitz,
\newblock ``Portfolio selection,''
\newblock {\em The Journal of Finance}, vol. 7, no. 1, pp. 77--91, 1952.

\bibitem{zeyde2012single}
Roman Zeyde, Michael Elad, and Matan Protter,
\newblock ``On single image scale-up using sparse-representations,''
\newblock in {\em Curves and Surfaces: 7th International Conference, Avignon, France, June 24-30, 2010, Revised Selected Papers 7}. Springer, 2012, pp. 711--730.

\bibitem{bevilacqua2012low}
Marco Bevilacqua, Aline Roumy, Christine Guillemot, and Marie~Line Alberi-Morel,
\newblock ``Low-complexity single-image super-resolution based on nonnegative neighbor embedding,''
\newblock 2012.

\bibitem{Agustsson_2017_CVPR_Workshops}
Eirikur Agustsson and Radu Timofte,
\newblock ``Ntire 2017 challenge on single image super-resolution: Dataset and study,''
\newblock in {\em The IEEE Conference on Computer Vision and Pattern Recognition (CVPR) Workshops}.

\bibitem{zamir2022restormer}
Syed~Waqas Zamir, Aditya Arora, Salman Khan, Munawar Hayat, Fahad~Shahbaz Khan, and Ming-Hsuan Yang,
\newblock ``Restormer: Efficient transformer for high-resolution image restoration,''
\newblock in {\em Proceedings of the IEEE/CVF conference on computer vision and pattern recognition}, 2022, pp. 5728--5739.

\bibitem{lee2021iterative}
Junyong Lee, Hyeongseok Son, Jaesung Rim, Sunghyun Cho, and Seungyong Lee,
\newblock ``Iterative filter adaptive network for single image defocus deblurring,''
\newblock in {\em Proceedings of the IEEE/CVF Conference on Computer Vision and Pattern Recognition}, 2021, pp. 2034--2042.

\bibitem{chen2021learning}
Liang Chen, Jiawei Zhang, Jinshan Pan, Songnan Lin, Faming Fang, and Jimmy~S Ren,
\newblock ``Learning a non-blind deblurring network for night blurry images,''
\newblock in {\em Proceedings of the IEEE/CVF Conference on Computer Vision and Pattern Recognition}, 2021.

\bibitem{wang2023reverse}
Lizhong Wang, Pierre-Alain Fayolle, and Alexander~G Belyaev,
\newblock ``Reverse image filtering with clean and noisy filters,''
\newblock {\em Signal, Image and Video Processing}, vol. 17, no. 2, pp. 333--341, 2023.

\bibitem{mittal2012no}
Anish Mittal, Anush~Krishna Moorthy, and Alan~Conrad Bovik,
\newblock ``No-reference image quality assessment in the spatial domain,''
\newblock {\em IEEE Transactions on image processing}, vol. 21, no. 12, pp. 4695--4708, 2012.

\end{thebibliography}
\end{document}